\documentclass[aps,noshowpacs,amsmath,amssymb]{revtex4}

\usepackage{graphicx}
\usepackage{hyperref}

\begin{document}

\title{Double parton scattering as a source of quarkonia pairs in LHCb}

\author{Alexey Novoselov} \email[]{Alexey.Novoselov@cern.ch}
\affiliation{Institute for High Energy Physics, Protvino, Russia}
\affiliation{Moscow Institute of Physics and Technology, Dolgoprudny, Russia}


\begin{abstract}
Recent results on production of $J/\psi$ pairs in LHCb initiated discussion on double parton scattering (DPS)
contribution to this process, which is not yet elaborated well. In this short report $J/\psi \chi_c$ and
$J/\psi \Upsilon$ modes are proposed for DPS studies. Estimations for  $\Upsilon \Upsilon$ and $J/\psi \Upsilon$
production in LHCb are presented.
\end{abstract}


\maketitle


Recent observation of $J/\psi$-meson pairs at LHCb stimulated discussion on phenomena contributing this
process. On the one hand, there is leading order (LO) calculation of the $gg \to J/\psi J/\psi$ process 
in the color-singlet (CS) model \cite{Humpert:1983yj,Berezhnoy:2011xy,Qiao:2009kg,Ko:2010xy}. 
It leads to the total cross section at $7$ TeV energy of $24$ nb and 
to $4$ nb in LHCb conditions. On the other hand, high flux of incoming partons at LHC energies leads 
to significant probability of scattering of more than one parton pairs in the same proton-proton collision \cite{Berger:2011jn}. 
Estimations of this double parton scattering (DPS) contribution to the $J/\psi$ pairs production \cite{Kom:2011bd}
lead to the cross section of approximately $2$ nb in the LHCb acceptance \cite{Baranov:2011ch}. 
Both SPS and DPS estimations are of the order of first experimental measurement in LHCb \cite{belyaev}, 
$\sigma({pp \to J/\psi J/\psi+X})=5.6 \pm 1.1$ nb. Meanwhile, both these predictions have some uncertainties.
In SPS calculations they are induced mainly by dependencies on $\alpha_s$ and $m_c$ parameters. 
DPS estimations does not account for correlations in double parton density functions and depend on 
a phenomenological parameter $\sigma_{\rm eff}$. Thus further investigation is desirable.


In this work other di-quarkonia final states are considered in addition to the $J/\psi J/\psi$ one. 
It will be shown that for $J/\psi\chi_c$ and $J/\psi\Upsilon$ production DPS should be main source at least at low-invariant-mass region. 


Assuming factorization of two hard partonic processes $A$ and $B$ 
one can write inclusive cross section of a double parton 
scattering process in a hadron collision in the following form \cite{Paver:1982yp}:
\begin{eqnarray} 
\label{hardAB}
\sigma^{A B}_{\rm DPS} &=& \frac{m}{2} \sum \limits_{i,j,k,l} \int 
\Gamma_{ij}(x_1, x_2, {\bf b_1},{\bf b_2}, Q^2_1, Q^2_2) \times 
\hat{\sigma}^A_{ik}(x_1, x_1^{'},Q^2_1) 
\hat{\sigma}^B_{jl}(x_2, x_2^{'},Q^2_2)
\nonumber
\\
&\times& \Gamma_{kl}(x_1^{'}, x_2^{'}, {\bf b_1} - {\bf b},{\bf b_2} - {\bf b}, 
Q^2_1, Q^2_2)
\times dx_1 dx_2 dx_1^{'} dx_2^{'} d^2b_1 d^2b_2 d^2b,
\end{eqnarray}
where $\Gamma_{ij}(x_1, x_2;{\bf b_1},{\bf b_2}; Q^2_1, Q^2_2)$ is double parton 
distribution function (PDF), $\hat{\sigma}_{ik}^{A,B}$ --- partonic cross sections of processes in question,
${\bf b}$ --- impact parameter and $m=2$ for different partinic subprocesses and $1$ --- for identical. 
It is usually assumed that longitudinal and transverse components of the PDF can be decomposed in the following way:
\begin{eqnarray} 
\label{DxF}
\Gamma_{ij}(x_1, x_2;{\bf b_1},{\bf b_2}; Q^2_1, Q^2_2)
&=& D^{ij}_h(x_1, x_2; Q^2_1, Q^2_2) f({\bf b_1}) f({\bf b_2}),
\end{eqnarray} 
and longitudinal component $D^{ij}_h(x_1, x_2; Q^2_1, Q^2_2)$ is taken as a product of two independent 
single parton distributions,
\begin{eqnarray} 
\label{DxD}
D^{ij}_h(x_1, x_2; Q^2_1, Q^2_2) &=& D^i_h (x_1; Q^2_1) D^j_h (x_2; Q^2_2).
\end{eqnarray}
This leads to a well-known simple expression for DPS cross-section in which no di-parton correlations are involved:
\begin{eqnarray} 
\label{doubleAB}
& \sigma^{ A B }_{\rm DPS} = \frac{m}{2} \frac{\sigma^{ A}_{\rm SPS} 
\sigma^{ B}_{\rm SPS}} {\sigma_{\rm eff}}.
\end{eqnarray} 
CDF and D0 measurements give $\sigma_{\rm eff} = 14.5$ mb, which is 
roughly 30$\%$ of the non-single diffraction (NSD) cross section at the Tevatron ($\approx 48 {\rm mb}$). 
The NSD cross section at LHC is only slightly higher ($\approx 51 {\rm mb}$). This supports an assumption that
$\sigma_{\rm eff}$ weakly depends on the total energy of interaction \cite{flensburg}.
Nonetheless dependence on the resolution scale and consequently on the partonic process
can be significant. Model of multiple interactions implemented in Pythia8 \cite{Corke:2009pm} MC generator
tries to accounts for this dependence and predicts $\sigma_{\rm eff}$ for double charmonia production of 30 mb 
($\sigma_{\rm eff} = \sigma_{\rm NSD} / \langle f_{\rm impact} \rangle $). Further we will use usual
value of $14.5~{\rm mb}$.

LO calculations in the CS model \cite{Berezhnoy:2011xy} lead do the value of double prompt $J/\psi$ production cross section 
in LHCb acceptance ($2 < y < 4.5$) of approximately 
\begin{eqnarray} 
\sigma_{\rm SPS}^{pp\to J/\psi J/\psi+X} &=& 4 ~ {\rm nb}.
\end{eqnarray} 
This result depends on the $\psi_{J/\psi}(0)$, $m_c$ and $\alpha_s$ parameters. First of them, $\psi_{J/\psi}(0)$,
can be determined rather precisely from the leptonic width. $m_c$ was set equal to $m_{J/\psi}/2$.
LO running $\alpha_s$ at the transverse mass scale was used.
Taking experimentally measured cross section of single $J/\psi$ production in LHCb of $10 {\rm{\mu}b}$ \cite{psiLHCb} and using expression
(\ref{doubleAB}) one gets for the DPS contribution approximately
\begin{eqnarray} 
\sigma_{\rm DPS}^{pp\to J/\psi J/\psi+X} &=& 2 ~ {\rm nb}.
\end{eqnarray} 
Sum of these contributions agrees well with the experimentally measured value \cite{belyaev}
\begin{eqnarray} 
\sigma_{\rm exp.}^{pp\to J/\psi J/\psi+X} &=& 5.6 \pm 1.1 ~ {\rm nb}.
\end{eqnarray} 

Several methods to distinguish DPS contribution from the SPS one were proposed \cite{Kom:2011bd,Berger:2011jn}. However it would
be easer to deal with process in which one of these contributions in suppressed compared to another.

Crucial feature of the CS model is presence of so-called selection rules. According to the C-parity conservation,
C-parity of the final state must be the same as those of 2 initial gluons, which is C-even as they are in CS combination. 
That is why production of $J/\psi \chi_c$ in SPS is expected to be suppressed. DPS cross section has no suppression
in this mode and should be significant as it is known that about $50\%$ of single $J/\psi$ mesons originate from
the feeddown from the $\chi_c$ decays.
Thus observation of $J/\psi$-pairs accompanied by a photon (from the $\chi_c \to J/\psi \gamma$ decay) would be a 
signal of DPS contribution.

Let us consider $\Upsilon(1S)$-meson pair production. Calculations analogous to $J/\psi$-pair production in \cite{Berezhnoy:2011xy}
lead to the total cross section at $7$ TeV of $31$ pb and in LHCb acceptance: 
\begin{eqnarray} 
\sigma_{\rm SPS}^{pp\to\Upsilon\Upsilon+X} &=& 8.7 ~ {\rm pb},
\end{eqnarray} 
while estimation of DPS cross section based on the experimental data on single $\Upsilon$ production at LHC \cite{upsLHCb}
gives
\begin{eqnarray} 
\sigma_{\rm DPS}^{pp\to\Upsilon\Upsilon+X} &=& 0.4 ~ {\rm pb}.
\end{eqnarray} 
One sees that in the $\Upsilon(1S)\Upsilon(1S)$ mode SPS should dominate. Also feeddown from $\Upsilon(2S)$ and $\Upsilon(3S)$ decays
should increase SPS prediction while it is already accounted for in the DPS estimation.

\begin{figure}
\begin{centering}
\includegraphics[width=12cm]{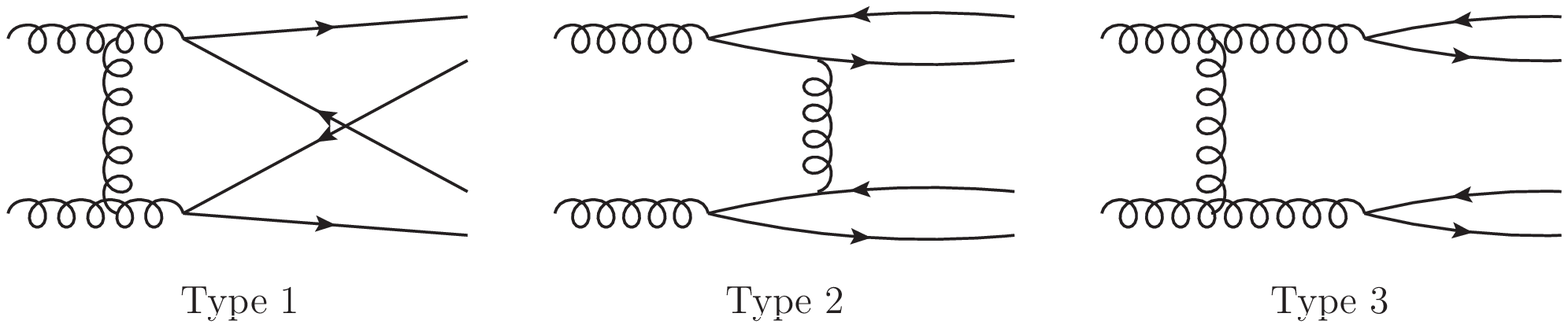} 
\end{centering}
\caption{Feynman diagrams for $gg \to q \bar{q} q \bar{q}$ process. \label{fig:diags}}
\end{figure}

There are basically $3$ types of Feynman diagrams for the $gg \to q \bar{q} q \bar{q}$ process (see fig. \ref{fig:diags}). 
Diagrams of first type contain $1$ fermion loop while diagrams of second type contain $2$ fermion loops connected by
an intermediate gluon. In the diagrams of second type final particles are coupled with $2$ gluons and consequently these diagrams
do not contribute to $S$-wave pairs production due to the $C$-parity conservation. Diagrams of the third type correspond to
gluon fragmentation and contribute to CO-states production only. Thereby only diagrams of first type contribute CS $J/\psi$ or $\Upsilon$ 
pair produced. But there are no LO diagrams contributing CS $J/\psi \Upsilon$ combined production. 
This final state is however accessible 
through $\chi_c \chi_b$ production followed by $\chi_c \to J/\psi + X$ and $\chi_b \to \Upsilon + X$ decays. Meanwhile $\chi_c \chi_b$
production does not exhibit kinematical peak near the threshold and is not expected to be significant in the low invariant mass region.
Prediction of $J/\psi \Upsilon$ production in CO mechanism made in \cite{Ko:2010xy} give in LHCb acceptance 
\begin{eqnarray} 
\sigma_{\rm SPS}^{pp\to J/\psi_8 \Upsilon_8+X} &=& 2 ~ {\rm pb},
\end{eqnarray}
while DPS contribution found using expression (\ref{doubleAB}) gives approximately 
\begin{eqnarray} 
\sigma_{\rm DPS}^{pp\to\Upsilon J/\psi+X} &=& 75 ~ {\rm pb}.
\end{eqnarray} 
So in the $J/\psi \Upsilon$ mode prediction for DPS exceeds significantly those for SPS one. It is interesting to notice that for SPS
$\sigma_{\rm SPS}^{J/\psi\Upsilon}<\sigma_{\rm SPS}^{\Upsilon\Upsilon}<\sigma_{\rm SPS}^{J\psi J\psi}$
while for DPS $\sigma_{\rm DPS}^{\Upsilon\Upsilon}<\sigma_{\rm DPS}^{J/\psi\Upsilon}<\sigma_{\rm DPS}^{J\psi J\psi}$.

Thereby cross section of $J/\psi J/\psi$ production is saturated by both SPS and DPS contributions, while
for $\Upsilon\Upsilon$ and $J/\psi\Upsilon$ final states only one of these regimes prevail. 

Author would like to thank A.K. Likhoded and A.V. Luchinsky for fruitful
discussions. The article was supported by Russian Foundation
for Basic Research (grant \#10-02-00061a), non-commercial foundation
``Dynasty'' and the grant of the president of Russian Federation (grant \#MK-406.2010.2).

\end{document}